\documentclass[12pt]{article}
\usepackage{amssymb}

\begin{document}

\noindent

\def\med{{1\ov 2}}
\def\hepth#1{ {\tt hep-th/#1}}

\def\be{\begin{equation}}
\def\ee{\end{equation}}
\def\bes{\begin{equation*}}
\def\ees{\end{equation*}}

\def\beqa{\begin{eqnarray}}
\def\beqas{\begin{eqnarray*}}
\def\eeqa{\end{eqnarray}}
\def\eeqas{\end{eqnarray*}}
\def\bea{\begin{eqnarray}}
\def\eea{\end{eqnarray}}

\def\cl{\mbox{\tiny (class)}}

\def\etal{{\it et al.\/}}%--- et al.
\def\ie{{\it i.e.\/}}%--- i.e.
\def\eg{{\it e.g.\/}}%--- e.g.
\def\Cf{{\it Cf.\ }}%--- Cf.
\def\cf{{\it cf.\ }}%--- cf.

\def\al{\alpha}
\def\e {\epsilon}
\def\lam{\lambda}
\def\blam{\bar\lambda}
\def\th{\theta}
\def\bth{\bar\theta}
\def\bsigma{\bar\sigma}
\def\bpsi{\bar\psi}

\def\bu{{\bar 1}}
\def\bd{{\bar 2}}
\def\bt{{\bar 3}}
\def\bc{{\bar 4}}
\def\dgam{\dot\gamma}
\def\dal{\dot\alpha}
\def\dbet{{\dot\beta}}

\def\Re{{\rm Re}}
\def\Im{{\rm Im}}

\def\H{{\cal H}}
\def\tr{{\rm tr}}
\def\Tr{{\rm Tr}}
\def\F{{\cal F}}
\def\N{{\cal N}}

\def\d{\partial}
\def\ov{\over}
\def\bD{\bar D}

\def\pder#1#2{{{\partial #1}\over{\partial #2}}}%--- partial derivative
\def\der#1#2{{{d #1}\over {d #2}}}%--- full derivative
\def\ppder#1#2#3{{\partial^2 #1\ov\partial #2\partial #3}}
\def\dpder#1#2{{\partial^2 #1\ov\partial #2 ^2 }}
\def\bemat{\left(\begin{array}}
\def\enmat{\end{array}\right)}
\def\theequation{\thesection.\arabic{equation}}

\def\Fpk{\alpha\!\cdot\!\!\F'_k}
\def\Fpu{\alpha\!\cdot\!\!\F'_1}
\def\Fpb{\beta\!\cdot\!\!\F'_1}
\def\Fpd{\alpha\!\cdot\!\!\F'_2}
\def\Fppk{\alpha\!\cdot\!\!\F''_k\!\!\cdot\! \alpha}
\def\Fppu{\alpha\!\cdot\!\!\F''_1\!\!\cdot\! \alpha}
\def\Fppb{\beta\!\cdot\!\!\F''_1\!\!\cdot\! \beta}
\def\Fppd{\alpha\!\cdot\!\!\F''_2\!\!\cdot\! \alpha}
\def\cdotsh{\!\cdot}
\def\cdotsk{\!\cdot\!}

%%%%%%%%%%%%%%%%%%%%%%%%%%%%%%%%%%%%%%%%%%%%%
%%%%%%%%%%%%%%%%%%%%%%%%%

\begin{titlepage}
\begin{flushright}
{ ~}\vskip -1in
US-FT/9-99\\
\hepth{9904087}\\
April 1999\\
\end{flushright}

\vspace*{20pt}
\bigskip

\centerline{\Large INSTANTON CORRECTIONS IN ${\cal N}=2$ SUPERSYMMETRIC}
\bigskip
\centerline{\Large THEORIES WITH CLASSICAL GAUGE GROUPS}
\bigskip
\centerline{\Large AND FUNDAMENTAL MATTER HYPERMULTIPLETS}
\vskip 1.3truecm
\centerline{\large\sc Jos\'e D. Edelstein\footnote{\tt edels@fpaxp1.usc.es},
Marta G\'omez--Reino\footnote{\tt marta@fpaxp1.usc.es} and Javier
Mas\footnote{\tt jamas@fpaxp1.usc.es}}

\vspace{1pc}

\begin{center}
{\em Departamento de F\'\i sica de Part\'\i culas, Universidade de Santiago
de Compostela,\\
E-15706 Santiago de Compostela, Spain.} \\

\vspace{5pc}

{\large \bf Abstract}

\end{center}

We compute instanton corrections to the low energy effective prepotential
of ${\cal N}=2$ supersymmetric theories in a variety of cases, including all
classical gauge groups and even number of fundamental matter hypermultiplets.
To this end, we take profit of a set of first- and second-order equations
for the logarithmic derivatives of the prepotential with respect to the
dynamical scale expressed in terms of Riemann's theta-function.
These equations emerge in the context of the Whitham hierarchy approach
to the low-energy Seiberg--Witten solution of supersymmetric gauge theories. 
Our procedure is recursive and allows to compute the effective prepotential to
arbitrary order in a remarkably straightforward way. General expressions for
up to three-instanton corrections are given. We illustrate the method with
explicit expressions for several cases.

\end{titlepage}
\setcounter{footnote}{0}

\section{Introduction}

Some five years ago, Seiberg and Witten gave an ansatz for the dominant piece
of the effective action governing the light degrees of freedom of $SU(2)$
${\cal N} = 2$ super Yang-Mills theory at low energy\cite{SeiWitt}. It is
given in terms of an auxiliary complex algebraic curve ${\cal C}$ (whose
moduli space is identified with the quantum moduli space of the low-energy
theory ${\cal M}_\Lambda$) and a given meromorphic differential,
$dS_{SW}$, that induces a special geometry on ${\cal M}_\Lambda$ (see
Refs.\cite{reviews} for excellent reviews).
The solution was soon extended to other gauge groups and matter content by
determining both the appropriate complex curve and meromorphic
differential \cite{sun,klemm,sunmat,hananoz,son}, thus leading to a
substantial progress in our understanding of ${\cal N}=2$ supersymmetric
gauge theories.
In particular, the appearance of an auxiliary Riemann surface made it possible
to identify remarkable connections with string theory, singularity theory of
differentiable maps and integrable systems. The framework that will be used
in this paper is strongly inspired by the latter one.

It is well-known that, as long as  ${\cal N}=2$ supersymmetry is
unbroken, the low energy effective action is given in terms of a holomorphic
prepotential $\F$.
The solution proposed by Seiberg and Witten embodies a prescription to
compute this prepotential.
However, its explicit evaluation for a given gauge group and matter content
is technically involved and it requires
to integrate an expression for the B-periods of $dS_{SW}$ as functions of its
A-periods. The complexity of this procedure increases rapidly with the rank of
the gauge group, even without matter hypermultiplets.

Whereas perturbative contributions to $\F$ are exhausted by one-loop diagrams
\cite{Sei}, the non-perturbative part is given by an infinite
series of instanton corrections.
The importance of instanton calculus lies precisely in the fact that it
provides one of the few non-perturbative links between the Seiberg--Witten
solution and the microscopic non-abelian field theory that it is supposed to
describe effectively at low energies.  From the microscopic point of view, the
first few instanton contributions to the asymptotic semiclassical expansion
of the effective prepotential have been computed for gauge group $SU(N_c)$,
and a remarkable agreement with the Seiberg--Witten solution has been found
\cite{hunter}.  From the side of the effective theory, several methods for
determining the instanton corrections have been developed in the last few
years by using the Picard--Fuchs equations \cite{klemm,itoyang1},
holomorphicity arguments \cite{itosas1}, analytic continuation \cite{dhoker}
(also for non-hyperelliptic curves \cite{isa}), modular anomaly equations
\cite{MNW}, etc. Among them, we would like to distinguish those methods that
lead to recursion relations for the $k$-instanton corrections, as long as they
give an implicit expression for the {\em exact} solution.
For the case of $SU(2)$, recursion relations determining the whole instanton
expansion have been obtained both in the pure gauge theory \cite{Matone} and
when matter is included \cite{itoyang1}. These recursion relations were
obtained by combining the renormalization group and Picard-Fuchs equations.
Also one-instanton corrections for the whole ADE series where obtained along
the same lines in \cite{itoyang}. 
$SU(N_c)$ with additonal matter in the adjoint representation was considered
in \cite{DHPh} from the point of view of the Calogero-Moser model, and in
\cite{MNW} (for ADE groups with dual Coxeter number $k_D\leq 6$) where the
modular anomaly equations of softly broken ${\cal N}=4$ supersymmetric gauge
theories were invoked. 

In a recent paper, \cite{emm}, a new strategy was observed to work
very well for the case of pure $SU(N_c)$.
It is based on a set of first- and second-order equations for the logarithmic
derivatives of the prepotential with respect to the dynamical scale $\Lambda$,
evaluated all over the moduli space
\beqa
\pder{\F}{\log\Lambda\,} & = & {\beta\ov 2\pi i } u_{2}~,
\label{lder} \\
\dpder{\F}{(\log\Lambda)} & = & -{\beta^2\ov 2\pi i}
\pder{u_{2}}{ a^i}\pder{u_{2}}{ a^j}{1\ov i\pi}
\d_{\tau_{ij}}\log\Theta_E(0|\tau) ~, \label{llder}
\eeqa
where the different quantities entering these expressions will be explained
below. Each one of these equations was obtained separately in the last few
years by many authors \cite{Matone,marimoore,losev,EY,rge}. It was not until
very recently that a unifying approach based on the Whitham hierarchy was
shown to be useful to obtain both \cite{ITEP}. In Ref.\cite{emm}, the ansatz
for the semiclassical expansion of the prepotential for gauge group
$SU(N_c)$ was inserted in both sides of equations (\ref{lder})--(\ref{llder})
with the result of an elegant and systematic procedure that allowed us to
compute instanton corrections {\em up to any desired order} with relatively
little effort. In particular, this method does not require knowledge of the
actual solution for the periods $a$ and $a_D$ of $dS_{SW}$, a fact which
spares a considerable amount of work.

In this paper we will exhibit the strengh of this method by extending the
results of \cite{emm} to any classical gauge group with and without matter
content. We shall limit ourselves to asymptotically free theories. The
obvious question arises, about the validity of equations like (\ref{llder})
for all these situations.
It turns out that, for our present purposes, the only constraint which seems
to be unavoidable within this method is that massive hypermultiplets have to
be introduced in pairs, degenerated in mass.
This can be understood either from a purely theoretical study, which we shall
leave for a separate paper \cite{egrmm}, or else, ``a posteriori", for the
consistency of the results.

Our aim is that this paper could be useful to anyone interested in finding
explicit expressions for the instanton corrections to the effective
prepotential of ${\cal N}=2$ supersymmetric gauge theories. To this end,
general formulas for up to three-instanton corrections will be given and the
method to obtain arbitrary higher corrections will be clearly explained. A
number of particular examples will be also worked out for the lower rank
groups and small number of flavour-pairs, in order to make them easily
available for futher comparison even to other methods or, actually, to the
results obtained in the microscopic non-Abelian field theory.
For the one- and two-instanton corrections, our results coincide with those
in the literature while, for most of the three-instanton contributions, our
results are new.

%%%%%%%%%%%%%%%%%%%%%%%%%%%%%%%%%%%%%%%%%%%%%%%%%%%%%%%%%%%%%%%%%%%%%%%%%%%%%
%%%%%%%%%%%%%%%%%%%%%%%%%%%%%%%%%%%%%%%%%%%%%%%%%%%%%%%%%%%%%%%%%%%%%%%%%%%%%
%%%%%%%%%%%%%%%%%%%%%%%%%%%%%%%%%%%%%%%%%%%%%%%%%%%%%%%%%%%%%%%%%%%%%%%%%%%%%

\section{Recursive Evaluation of the Effective Prepotential}
\setcounter{equation}{0}
\indent

\subsection{ Review and Notation}

For completeness, we shall start by reviewing the r\^ ole of the different
ingredients that enter the formulas (\ref{lder}) and (\ref{llder}).
The low-energy dynamics of ${\cal N}=2$ supersymmetric theories with
classical gauge group $G$ corresponding to $N_c$ colors, and $N_f$
hypermultiplets of mass $m_f = m_{(f+ N_f/2)}$ ({\it i.e.} degenerated in
masses)  in the fundamental representation of $G$,
can be described in terms of an auxiliary hyperelliptic curve ${\cal C}$ given
by
\be
y^2 = (P(\lambda,e_p)+T(\lambda,m_f,\Lambda))^2 -
4\Lambda^{\beta} F(\lambda,m_f) ~,
\label{curveone}
\ee
where $P$ is the characteristic polynomial of $G$, $\Lambda$ is the quantum
generated dynamical scale, $\beta$ is the coefficient of the one-loop ${\cal
N}=2$ beta function and $e_p$ are the eigenvalues of the complex scalar field
$\langle\phi\rangle = \sum_p e_p h_p$ (see Ref.\cite{Lie} for the conventions
followed in the notation of Lie group and Lie algebra objects) that belongs to
the ${\cal N}=2$ vector supermultiplet in the adjoint of $G$.
$T$ and $F$ are polynomials that do not depend on the {\em moduli} $e_p$ and
$T$ is different from zero only when $N_f>N_c$. 
As pointed out in Ref.\cite{dhoker}, when the gauge group is $SU(N_c)$ or
$SO(N_c)$ all dependence on $T$ can be absorbed in a redefinition of 
$e_p$, the effective prepotential remaining untouched. Thus,
we can set $T=0$, and write the hyperelliptic curve ${\cal C}$ as
\be
y^2 = P^2(\lambda,e_p) - 4\Lambda^{\beta} F(\lambda,m_f) ~.
\label{curve}
\ee
In the case of $Sp(N_c)$, there is a residual value for $T$, $T =
\Lambda^{N_c-N_f+2}(\prod_{f=1}^{N_f/2}m_f^2)$. In order to consider all the
different cases within a unified framework we will neglect this contribution
by setting the mass of one of the degenerated hypermultiplets to zero (say for
example $m_{N_f/ 2}=0$). Following Ref.\cite{dhoker} this case will be
denoted  $Sp(N_c)''$, and the corresponding hyperelliptic curve has the form
(\ref{curve}) with the proviso that, according to our previous remark, the
number of hypermultiplets will be at least two, $N_f=N_f'+2\geq 2$.
Consequently, the case of pure $Sp(N_c)$ will not be attainable from our
results though it is quite simple to make the appropriate modifications, as
will be explained below. It is convenient, for later use, to list the form
of $P$, $F$ and $\beta$ for all classical groups:
\vskip7mm
\centerline
{\begin{tabular}{|c||c|c|c|c|c|c|c|c|}
\hline
$G$ & $P(\lambda,e_p)$ & $F(\lambda)$ & $\beta/2$ & $l_G$ 
& $\xi$  & $\xi^+$ &
$\xi^-$ & $k_D$
  \\
\hline\hline
$SU(r+1)$  & $\displaystyle\prod_{p=1}^{r+1}(\lambda-e_p)$ &
$\displaystyle\prod_{f=1}^{N_f}(\lambda+m_f)$ & $r+1-N_f/2$ & r+1 & 1 & 1 & 0
& r+1
\\ \hline
$SO(2r)$   & $\displaystyle\prod_{p=1}^{r}(\lambda^2-e_p^2)$ &
$\displaystyle\lambda^4\prod_{f=1}^{N_f}(\lambda^2-m_f^2)$ & $ 2r-N_f-2$ & r
& 1 & 1 & 1 & 2r-2
\\ \hline
$SO(2r+1)$ & $\displaystyle\prod_{p=1}^{r}(\lambda^2-e_p^2)$ &
$\displaystyle\lambda^2\prod_{f=1}^{N_f}(\lambda^2-m_f^2)$ & $2r-N_f-1$  & r &
1 & 1 & 1 & 2r-1
\\ \hline
$Sp(2r)''$   & $\displaystyle\prod_{p=1}^{r}(\lambda^2-e_p^2)$ &
$\displaystyle\prod_{f=1}^{N_f'}(\lambda^2-m_f^2)$ & $2r-N_f'$  & r & 2 & 2
& 2 & r+1
\\ \hline
\end{tabular}}
\vskip5mm
\centerline{ {\em Table 1}}
\vskip7mm\noindent
The symmetric polynomials $\bar u_k(e_p)$ and $ t_i(m_f)$ are defined through
the expansions
\beqa
P(\lambda,e_p) &\equiv  & \lambda^{h} - \sum_{k=2}^{h} \lambda^{h-k}
\bar u_k(e_p) ~,
\label{casimirs} \\
F(\lambda,m_f) &\equiv &\lambda^{h} + \sum_{k=1}^{h} \lambda^{h-k}
t_k(m_f) ~, \label{lastes}
\eeqa
where $h$ stands, in each case, for the highest power in $P(\lambda,e_p)$ or
$F(\lambda,m_f)$. The moduli for each curve (\ref{curve}) can be taken to be
either the independent roots $e_p$ or the  $a^i$ defined as the coefficients
of $\langle \phi\rangle = a^i H_i$ in  the Chevalley basis, hence  linearly
related to $e_p$ (see Ref.\cite{Lie}). Neither of these parameters are
invariant under Weyl transformations which, in particular,  act by permutation
on the $e_p$. On the contrary, the symmetric polynomials (Casimirs) $\bar u_k$
provide faithfull coordinates for the moduli space of vacua.
In particular, $\bar u_2 ={1\ov 2}\tr \phi^2=  (\frac{\xi^+
+\xi^-}{2\xi^+})\sum_{p=1}^{l_G} e_p^2 $, and in general:
$\bar u_k = {1\ov k} \tr \phi^k+$ ..., the dots standing for homogeneous
powers of lower Casimir operators.

The Seiberg--Witten meromorphic differential can be written as
\be
dS_{SW}(u_k,m_f) = \left({P}' - \frac{P F'}{2 F}\right)
{\lambda d\lambda \ov y} ~,
\label{diff}
\ee
and the quantum relations between the low-energy coordinates of the
moduli space $a^i, a_{D\,j}$ and the ``mean field" order parameters $u_k(a) =
{1\ov k} \tr \langle
\phi^k\rangle + \cdots =  \bar u_k(a^i) + {\cal O}(\Lambda)$ are implicitely
given by the period integrals
\be
a^i(u_k,m_f) = \oint_{A^i} dS_{SW}(u_k,m_f) ~~~~;~~~~
a_{D\,j}(u_k,m_f) = \oint_{B_j} dS_{SW}(u_k,m_f) ~,
\label{theper}
\ee
where $A^i$ and $B_j$ constitute a symplectic basis of homology cycles with
canonical intersections of the hyperelliptic curve (\ref{curve}); 
the effective prepotential $\F$ is implicitely defined by the equation
\be
a_{D\,i} = \pder{\F(a)}{a^i} ~,
\label{implicit}
\ee
so that its exact determination involves the integration of
functions $a_{D\,i}(a)$ for which there is not a closed form available.
In this context, the existence of an algorithm that let us determine the
exact form of $\F$ without going through the actual computation of
$a^i(u_k,m_f)$ and $a_{D\,j}(u_k,m_f)$ is welcome.

As mentioned above, our first ingredient is the set of RG equations
(\ref{lder})--(\ref{llder}). Strictly speaking, as they stand, they are valid
for the pure gauge theory. In presence of matter, the first equation receives
an additional term which depends only on the masses 
and the second one has to be modified
 for $\beta=2$ where it receives an additional constant contribution
\cite{egrmm}. In summary\footnote{
Although the addition of an $a^i$-independent terms to $\F$ is unphysical from
the point of view of the effective theory, the embedding of the Seiberg-Witten
solution into the Whitham dynamics fixes them as a function of the bare
coupling $\tau_0$. A similar behaviour  was observed in the study of
$\F$ near the strong coupling singularities of $SU(N_c)$
$\N=2$ super Yang--Mills theory \cite{edemas}.}
\beqa
\pder{\F}{\log\Lambda\,} & = & {\beta\ov 2\pi i } u_{2} + \varphi(m)~,
\label{lderm} \\
\dpder{\F}{(\log\Lambda)\,} & = &
-{\beta^2\ov 2\pi i} \pder{u_2}{a^i} \pder{u_2}{a^j} {1\ov
i\pi}\d_{\tau_{ij}}\log\Theta_E(0|\tau) + {\beta^2\ov 2\pi i}\Lambda^2
\delta_{\beta,2} ~.
\label{formula}
\eeqa
In Eq.(\ref{formula}),
$\Theta_E(0|\tau)$ is Riemann's theta function associated to the hyperelliptic
curve ${\cal C}$
\be
\Theta \left[ {\vec\alpha\atop \vec\beta}\right]
(\xi|\tau)=
\sum_{n_k \in {\bf Z} }\displaystyle{ e^{i
\pi
\bigl[  \tau_{ij}(n_i+ \alpha_i)
(n_j+ \alpha_j) + 2 (n_i+ \alpha_i)(\xi + \beta_i) \bigr]} }~,
\label{thetaf}
\ee
where $E = \left[ {\vec\alpha\atop \vec\beta}\right]$ stands for an even half
integer characteristic. In almost all the cases this characteristic will be
$E=\left[{0,\dots,0\atop \med,\dots,\med}\right]$, as it is in the case of
pure $SU(N_c)$ \cite{emm}.
The only exception is $Sp(2r)''$ for which the characteristic gets modified to
the value $E=\left[{0,\dots,0,0\atop 0,\dots,0,\med}\right]$.

The second ingredient is an ansatz for the instanton expansion of the
prepotential valid for any classical gauge group $G$ and $N_f$ massive
hypermultiplets with paired masses
\bea
\F_{G,N_f} & = & \frac{\tau_0^G}{4\pi i}\sum_{\alpha_+} Z_{\alpha_+}^2
+ {i\ov 4 \pi}\xi \sum_{\alpha_+} Z_{\alpha_+}^2 \log \,{Z_{\alpha_+}^2\ov
\Lambda^2}-{i\ov 4\pi}\xi^+
\sum_{p=1}^{l_G}\sum_{f=1}^{N_f/2}(e_p+m_f)^2\log\,{{(e_p+m_f)^2}\ov
{\Lambda^2}} \nonumber\\
&& -{i\ov 4\pi}\xi^-
\sum_{p=1}^{l_G}\sum_{f=1}^{N_f/2}(e_p-m_f)^2\log\,{{(e_p-m_f)^2}\ov
{\Lambda^2}}+{1\ov 2 \pi i}\sum_{k=1}^\infty
\F_{k}(Z)\Lambda^{k\beta} ~,
\label{elprep}
\eea
where $\alpha_+$ denotes a positive root and $\sum_{\alpha_+}$ is the sum
over all positive roots.
The set $\{ \alpha_i \}_{i=1,...,r}$ stands for the simple roots of the
corresponding classical Lie algebra, they generate the root lattice $\Delta =
\{\alpha = n^i \alpha_i| n^i\in {\mathbb{Z}}\}$. 
The dot product $(\cdot)$ of two simple roots $\alpha_i$ and $\alpha_j$ gives
an element of the Cartan matrix, $A_{ij}=\alpha_i\cdotsk\alpha_j$,
and extends bilinearly to arbitrary linear combinations of simple roots.
So, for example, for any root  $\alpha = n^j\alpha_j\in\Delta$,
the quantities $Z_\alpha$ are defined  by $Z_\alpha =
a\cdotsk \alpha \equiv a^i A_{ij} n^j$ where $a= a^i\alpha_i$. For non-simply
laced Lie algebras this product is not symmetric.

Simple roots can be written in terms of the  orthogonal set of unit vectors
$\{\e_p\}_{p=1,\cdots,l_G}$. The order parameters $a^i$ and $e_p$ are related
by $e_p = a\cdotsk\e_p$. The exact relations and the actual values of
$\alpha_i\cdotsk\e_p$ for each classical gauge group can be found in
Ref.\cite{Lie}. Finally, we also have that $\lambda^k \cdotsk \alpha_j=
\delta^k{_j}$ define the fundamental weights. In particular, this means that
$\alpha_i = \sum_k A_{ik}\lambda^k$. We have also introduced three parameters
$\xi,~\xi^+$ and $\xi^-$, so as to deal with all classical Lie algebras within
one single ansatz; $k_D$ denotes the dual Coxeter number. The particular
values of these variables for each classical gauge group are shown in Table 1.
The coefficient of the one-loop beta function turns out to be given by 
\be
\beta/2 = \xi  k_D -  \frac{N_f}{2}\left( 1 ~+~ \frac{\xi^-}{\xi^+}\right) ~.
\label{bfun}
\ee
By expressing the roots
$\alpha_+$ in terms of $\e_p$, ({\it i.e.}, $Z_{\alpha_+}$ in
terms of $e_p$) one can also check that
the following relation holds
\be
\xi^+~\sum_{p=1}^{l_G} e_p^2  = {1\ov k_D}~\sum_{\alpha_+}
Z_{\alpha_+}^2 ~.
\label{rel}
\ee

\subsection{The procedure}

It is important to notice that  we may shift $\tau_0^G$ in (\ref{elprep}) to
any value by appropriately rescaling
$\Lambda$, and this will be reflected in our choice for the normalization of
the $\F_{k}(Z)$.
We have fixed it  in all cases to be $\tau_0^G=3\beta/2k_D$ so that quadratic
terms in $Z_\alpha$ do not contribute to the coupling constant $\tau_{ij}$ (see
(\ref{latau}) below).

The l.h.s. of Eq.(\ref{formula}) can be easily computed from the expansion of
the effective prepotential (\ref{elprep}) to be
\be
\dpder{\F}{(\log \Lambda)} = {1\ov 2\pi i}
\sum_{k=1}^\infty (k \beta)^2~\F_k(Z) ~\Lambda^{k\beta} ~,
\label{makdj}
\ee
then comparing (\ref{makdj}) and (\ref{formula}) we get
\be
\sum_{k=1}^{\infty}k^2{\cal F}_k\Lambda^{k \beta}=-\pder{u_2}{a^i} \pder{u_2}
{a^j} {1\ov i\pi}\d_{\tau_{ij}}\log \Theta_E(0|\tau)~\label{ic}
+ \Lambda^2 \delta_{\beta,2} ~,
\ee
such that the instanton correction ${\cal F}_k$ can be obtained through a set
of recursive relations after expanding the r.h.s. of Eq.(\ref{ic}) in powers
of $\Lambda^{\beta}$.

The expansion of the derivative of the quadratic Casimir in powers of $\Lambda$
can be obtained from the RG equation (\ref{lderm}) and (\ref{elprep}),
\beqa
\pder{u_2}{a^i} &=& {2\pi i\ov \beta} \ppder{\F}{a^i}{\log\Lambda\,} \cr &=&
{2\xi\ov \beta}\sum_{\alpha_+}Z_{\alpha_+}\partial_iZ_{\alpha_+}
-{N_f\ov\beta } (\xi^++\xi^-)
\sum_p e_p\partial_i e_p+ \sum_{k=1}^\infty k
\F_{k,i} ~ \Lambda^{k \beta}\cr
&=& \xi^+\sum_pe_p\partial_ie_p + \sum_{k=1}^\infty k \F_{k,i} ~
\Lambda^{k \beta}\equiv
\sum_{k=0}^{\infty}{\H_{2,i}^{(k)}}\Lambda^{k \beta} ~,
\label{expanh}
\eeqa
where $\F_{k,i}= \d\F_k/\d a^i $ and use has been made of (\ref{bfun}) and
(\ref{rel}).

To expand the theta function we need to compute from
(\ref{elprep}) the couplings in the semiclassical region,
\beqa
\tau_{ij} &=&\ppder{\F}{a^i}{a^j} \nonumber \\
&=& {i\ov 2\pi}\xi \sum_{\alpha_+} \pder{Z_{\alpha_+}}{a^i}
\pder{Z_{\alpha_+}}{a^j} \log \left({ Z^2_{\alpha_+} \ov \Lambda^2} \right)
-{i\ov 2\pi}\xi^+
\sum_{p=1}^{l_G}\sum_{f=1}^{N_f/2}\pder{e_p}{a_i}\pder{e_p}{a_j}
\log~{{(e_p+m_f)^2}
\ov {\Lambda^2}}\nonumber\\
&&-{i\ov 2\pi}\xi^-
\sum_{p=1}^{l_G}\sum_{f=1}^{N_f/2}\pder{e_p}{a_i}\pder{e_p}{a_j}
\log~{{(e_p-m_f)^2} \ov {\Lambda^2}}+{1\ov 2\pi i} \sum_{k=1}^\infty \F_{k,ij}
\Lambda^{k \beta} ~,
\label{latau}
\eeqa
with $\F_{k,ij}= \ppder{\F_k}{a^i}{a^j}$. So
the term involving the couplings that appear in the theta function $\Theta_E$
can be written as
\bea
i\pi \, n^i\tau_{ij} n^j &=& \sum_{\alpha_+} \log
\left({Z_{\alpha_+}\ov \Lambda}\right)^{-\xi(\alpha\,\cdotsh\alpha_+)^2}
+~\sum_{p,f}\log~\left(\frac{e_p+m_f}{\Lambda}\right)^{\xi^+(\alpha\,\cdotsh
\e_p)^2}+ \nonumber\\ && + ~\sum_{p,f}
\log~\left(\frac{e_p-m_f}{\Lambda}\right)^{\xi^-(\alpha\,\cdotsh\e_p)^2} +~
\med\sum_{k=1}^\infty~ (\Fppk) ~\Lambda^{k \beta} ~,
\label{tauroot}
\eea
where $\Fppk\equiv\sum_{i,j}n^i\F_{k,ij} n^j$. 
Also  $\alpha= n^i\alpha_i$, and we have set
\be
n^i \left(\pder{Z_{\alpha_+}}{a^i}\right) = n^i~(\alpha_i\cdotsk\alpha_+) =
\alpha\cdotsk\alpha_+ ~~~~~;~~~~~~~ n^i \left(\pder{e_p}{a^i}\right) =
n^i~(\alpha_i\cdotsk\e_p) = \alpha\cdot\e_p ~.
\ee
Inserting (\ref{tauroot}) in the Theta function (\ref{thetaf}) with a
characteristic $E=\left[{0,\dots,0\atop \med,\dots,\med}\right]$, we obtain
\beqa
\Theta_E(0|\tau) &=& \sum_{\vec n} \exp\biggl[ i\pi n^i\tau_{ij} n^j + i \pi
\sum_{k} n_k\biggr]
\cr &=&
\sum_{\alpha\in\Delta}(-1)^{\rho\cdot \alpha} \prod_{\alpha_+}\left(
{Z_{\alpha_+}\ov \Lambda}\right)^{-\xi(\alpha\,\cdotsh\alpha_+)^2}
\prod_{p,f}\left({{e_p+m_f}\ov{\Lambda}}\right)^{\xi^+(\alpha\,\cdotsh\e_p)^2}
\prod_{p,f}\left({{e_p-m_f}\ov{\Lambda}}\right)^{\xi^-(\alpha\,\cdotsh\e_p)^2}
\cr  &&~~~~~~~\times\prod_{k=1}^\infty\exp\left({\med (\Fppk) \Lambda^{k
\beta}}\right)
\cr &=&
\sum_{s=0}^\infty \sum_{\alpha\in\Delta_s} (-1)^{\rho\cdot\alpha}
\prod_{\alpha_+} Z_{\alpha_+}^{-\xi(\alpha\,\cdotsh\alpha_+)^2}
\prod_{p}\left[ R(e_p)\right]^{(\alpha\,\cdotsh\e_p)^2} \nonumber\\
&&~~~~~~~\times\prod_{k=1}^\infty\left(\sum_{m=0}^\infty {1\ov 2^m
m!}\left(\Fppk\right)^m\, \Lambda^{k\beta m }\right) \Lambda^{s\beta }
\cr &\equiv&
\sum_{l=0}^\infty \Theta^{(l)} \Lambda^{l \beta} ~.
\label{expantheta}
\eeqa
In the previous expression, $\rho$ is the maximal weight $\rho = \sum_{i=1}^{r}
\lambda^i$. In the case of $Sp(N_c)''$,  due to its peculiar characteristic,
the dot product ${\rho\cdot \alpha}$ needs to be replaced by~
${\rho_r\cdotsh\alpha= n^r}$.  On the other hand, $\Delta_s\subset\Delta$ is a
subset of the root lattice composed of those lattice vectors
$\alpha\in\Delta_s$ that fulfill the constraint
$
\frac{1}{2}\xi^+ \sum_p(\alpha\cdotsh \e_p)^2 = s  ~.
$
The Weyl group permutes the $\e_p$, hence the previous statement is Weyl
invariant. In other words,  the sets $\Delta_s$ are unions of Weyl orbits.
This fact guarantees that the final result will recombine into Weyl invariant
expressions. In (\ref{expantheta}) we  have also introduced the polynomial
\beqa
R(\lambda,m_f)  &=& \prod_{f=1}^{N_f/2}(\lambda+m_f)^{\xi^+}
\prod_{f=1}^{N_f/2}(\lambda-m_f)^{\xi^-} \label{erres}\\
&\equiv& \left(\lambda^h + \sum_{i=1}^{h}
q_i(m_f)\lambda^{h-i}\right)^{\xi^+} ~,
\label{lasquses}
\eeqa
where $h$, again, stands for the appropriate highest power. It is at this point
where the need for pairwise equal masses enters. Otherwise, we would be
dealing with square root factors of the form  $(\lambda
\pm m_f)^{1/2}$. Now we can collect the first few terms in
the expansion (\ref{expantheta}),
\[ \Theta^{(0)} = 1 ~, ~~~~~~~~~~
\Theta^{(1)} = \sum_{\alpha\in\Delta_1}(-1)^{\rho\cdot\alpha}
\prod_{\alpha^+} Z_{\alpha_+}^{-\xi(\alpha\cdot\alpha_+)^2}
\prod_p\left[R(e_p)\right]^{(\alpha\cdot \e_p)^2} ~, \]
\beqa
\Theta^{(2)} & = & \sum_{\alpha\in\Delta_1}(-1)^{\rho\cdot\alpha}\,\med
(\Fppu ) \prod_{\alpha^+} Z_{\alpha_+}^{-\xi(\alpha\cdot\alpha_+)^2}
\prod_p\left[R(e_p)\right]^{(\alpha\cdot \e_p)^2} \nonumber\\
& & + \sum_{\beta\in\Delta_2}(-1)^{\rho\cdot\beta} \prod_{\alpha^+}
Z_{\alpha_+}^{-\xi(\beta\cdot\alpha_+)^2}
\prod_p\left[R(e_p)\right]^{(\alpha\cdot \e_p)^2} \nonumber ~.
\eeqa
However, in the logarithmic derivative, the Theta function appears in the
denominator so we shall need the expansion of $\Theta(0|\tau)^{-1}$ in terms of
$\Lambda$. We can write this expansion as
\be
\Theta(0|\tau)^{-1} = \sum_{l=0}^{\infty} \Xi^{(l)}(\Theta) \,
\Lambda^{2Nl} ~.
\label{tauinv}
\ee
Here $\Xi^{(0)}(\Theta) = 1$ and for $\Xi^{(l)}(\Theta)$ we can write
in general
\be
\Xi^{(l)}(\Theta) = \sum_{(p_1,...,p_k) \in {\bf N}^k}^{p_1 + 2p_2 + ...
+ kp_k = l} \chi_{(p_1,...,p_k)} \prod_{i=1}^l (\Theta^{(i)})^{p_i} ~,
\label{tauinvcoef}
\ee
where the coefficients $\chi$ are parametrized by the
partition elements $(p_1,...,p_k)$. The first few values for these
parameters are, for example,
\[ \chi_{(1)} = -1 ~, ~~~ \chi_{(2,0)} = 1 ~, ~~~ \chi_{(0,1)} = -1 ~, ~~~
\chi_{(3,0,0)} = -1 ~, ~~~ \chi_{(1,1,0)} = 2 ~, ~~~ \chi_{(0,0,1)} = -1 ~, \]
and using these values we can immediately obtain the lower $\Xi^{(l)}(\Theta)$.

Next, we compute the derivative of
the theta function with respect to the period matrix 
\beqa
% & & \!\!\!\!\!\!\!\!\!\!\!\!\!\!\!\!\!\!\!\!\!\!\!\!
{1\ov i\pi}\d_{\tau_{ij}}\Theta_E(0,\tau) & = & \sum_{ n} n^i n^j
\exp\biggl[ i\pi n^k\tau_{kl} n^l + i \pi \sum_{k} n_k\biggr] \cr
& = & \sum_{s=1}^\infty \sum_{\alpha\in\Delta_s} (-1)^{\rho\cdot\alpha}
(\lambda^i\cdotsh\alpha)(\lambda^j\cdotsh\alpha)
\prod_{\alpha_+} Z_{\alpha_+}^{-\xi(\alpha\,\cdotsh\alpha_+)^2}
\prod_{p}\left[ R(e_p)\right]^{(\alpha\,\cdotsh\e_p)^2} \nonumber\\
& & ~~~~~~\times\prod_{k=1}^\infty\exp\left({\med (\Fppk)
\Lambda^{k\beta }}\right) \Lambda^{s \beta}\nonumber\\
&\equiv& \sum_{p=1}^\infty \Theta_{ij}^{(p)} \Lambda^{p\beta} ~.
\label{expanthij}
\eeqa
Now, collecting all the pieces and inserting them back into (\ref{formula}),
we find for $\F_k(Z)$ the following expression:
\be
\F_k(Z) = - k^{-2}
\sum_{p, q, l=0}^{p+q+l = k-1}\sum_{ij} {\cal H}_{2,i}^{(p)} {\cal
H}_{2,j}^{(q)} \Theta_{ij}^{(k-p-q-l)} \Xi^{(l)} ~,
\label{elresul}
\ee
in terms of previously defined coefficients.
If we look at the factors on the r.h.s. of Eq.(\ref{elresul}), it is easy
to see that they involve $\F_1,\F_2,...$ up to
$\F_{k-1}$. In fact, although both $\H_2^{(p)}$ and $\Theta^{(p)}$ depend
on $\F_1,....\F_p$, the indices within parenthesis reach at most the value
$k-1$ as $\Theta_{ij}^{(0)}=0$. Moreover $\Theta_{ij}^{(k)}$ depends on
$\F_1,...,\F_{k-1}$ since the vector $\alpha=0$ is missing from the lattice
sum. This ``lucky accident" has its origin in the particular form of
the characteristic in the semiclassical (duality) frame, and seems to be an
essential feature in order to build up a recursive procedure to {\em compute
all the instanton coefficients by starting just from the perturbative
contribution to} $\F(a)$ \cite{edemas}. For the first few cases we may develop
(\ref{elresul}) to find
\beqa
\F_1 &=& -
\sum_{ij} {\cal H}^{(0)}_{2,i} {\cal H}^{(0)}_{2,j} \Theta_{ij}^{(1)} \nonumber
\\
\F_2 & = &  - \frac{1}{4} \sum_{ij}
\left(\Theta_{ij}^{(2)} {\cal H}_{2,i}^{(0)} {\cal H}_{2,j}^{(0)}
+ \Theta_{ij}^{(1)}(2 {\cal H}_{2,i}^{(1)}{\cal H}_{2,j}^{(0)} -
{\cal H}_{2,i}^{(0)}{\cal H}_{2,j}^{(0)}\Theta^{(1)})\right) \nonumber\\
&\vdots& \nonumber
\eeqa
After some algebraic manipulations  (\ref{elresul})  admits the following
general form
\be
\F_k=-\frac{1}{k^2} \sum_{s=1}^k\sum_{ \alpha\in\Delta_{s}}
(-1)^{\rho\cdot\alpha}
\prod_{\alpha_+} Z_{\alpha_+}^{ -\xi (\alpha\,\cdotsh\alpha_+)^2}
\prod_{p}\left[R(e_p)\right]^{ (\alpha\,\cdotsh\e_p)^2}
\Phi_{k+1-s}(\alpha) ~,
\label{ff}
\ee
where the functions $\Phi_{k} (\alpha)$ depend on $\F_1\cdots, \F_{k-1}$ and
have to be evaluated case by case.  
For the first few we have
\beqa
\Phi_1(\alpha) &= & Z_{\alpha(G)}^2 ~, \label{phi0} \\
\Phi_2(\F_1,\alpha) & = & \F_1 + 2 (\Fpu ) Z_{\alpha(G)} + \med (\Fppu)
Z_{\alpha(G)}^2 ~, \label{phi1} \\
\Phi_3(\F_1,\F_2,\alpha) & = & 4\F_2 + 4 (\Fpd) Z_{\alpha(G)} + (\Fpu)^2
+ \med (\Fppu)\left( \F_1 + 2 (\Fpu) Z_{\alpha(G)} \right) \nonumber \\
& & + {1\ov 8} (\Fppu)^2 Z_{\alpha(G)}^2 + \med (\Fppd) \, Z_{\alpha(G)}^2 ~,
\label{phi2}
\eeqa
where $\alpha\!\cdot\!\!\F'_k = n^i\F_{k,i}$. 
Expressions (\ref{ff})-(\ref{phi2})  make patent the iterative character of
the procedure. ~$Z_{\alpha(G)}$ stands for $n^i \H_{2,i}^{(0)}$~, and for
simply laced groups, $Z_{\alpha(G)}=Z_\alpha$ while for non-simply laced, the
exact form will be given below. In the case $\beta =2$, as we can see from
Eq.(\ref{ic}) the first instanton correction acquires a shift
$\tilde{\F}_1=\F_1+1~.$ It soon becomes clear that, except for the simplest
cases, the concrete evaluation of the $\F_k$  has to be carried out by
symbolic computation. In the next sections we illustrate our procedure with
explicit examples in several cases  for the lower rank groups.

A last word concerning the possibility to split the masses is in order.
Generically, all resulting expressions involve powers of the degenerated masses
$\{m_f,~f=1,...,N_f/2\}$. They must be recovered from the exact result for
arbitrary masses in the coincidence limit $m_f = m_{f+N_f}$. The possibility
to go back unambiguously only happens for  low powers of $m_f$. As a thumb
rule, we have checked in several cases that the following prescription does
the job: for $SU(N_c)$ and when powers of $m_f$ are not higher than 2, 
$m_f\to {1\ov 2}(m_f + m_{f+N_f})$ and $m_f^2 \to m_f m_{f+N_f}$; while for the
rest of the groups and for (even) powers of $m_f$ not higher than 4,
$m_f^2\to {1\ov 2}(m_f^2 + m^2_{f+N_f})$ and $m_f^4 \to m^2_f m^2_{f+N_f}$.
Typically these cases only occur in $\F_1$. Another way to see this is to
observe that if we write down  $\F_1$ in terms of the $q_i(m_f),~ 
f=1,..N_f/2$ as given in (\ref{lasquses}), these factors always appear
precisely in those combinations that build up the $t_k= t_k(q_i)$ as given in
(\ref{lastes}), which are valid for arbitrary masses.

%%%%%%%%%%%%%%%%%%%%%%%%%%%%%%%%%%%%%%%%%%%%%%%%%%%%%%%%%%%%%%%%%%%%%%%%%%%%%
%%%%%%%%%%%%%%%%%%%%%%%%%%%%%%%%%%%%%%%%%%%%%%%%%%%%%%%%%%%%%%%%%%%%%%%%%%%%%
%%%%%%%%%%%%%%%%%%%%%%%%%%%%%%%%%%%%%%%%%%%%%%%%%%%%%%%%%%%%%%%%%%%%%%%%%%%%%

\section{Results for Simply-Laced Lie Algebras}
\setcounter{equation}{0}
\indent

We start by giving the concrete expression for $Z_{\alpha(G)} =
n^i\H_{2,i}^{(0)}$. In the case of $A_r$ and $D_r$ Lie algebras, we have that
$\sum_p e_p^2 = a^i a^j A_{ij} $ and also that $\xi^+=1$. Hence
$$
Z_{\alpha(G)} = n^i\H_{2,i}^0 =n^i\sum_p e_p\d_i e_p=n^i a^j A_{ij} = a\cdotsk
\alpha = Z_\alpha
$$
We shall define  $\Delta_0=\prod_{\alpha_+}Z_{\alpha+}^2~. $ To avoid
confusion we must mention that, although we have expressed all results in
terms of Weyl invariant polynomial combinations $\bar u_k(a^i)$, for
notational clearness we shall drop the bar.

\subsection{{\boldmath $SU(r+1)$} with {\boldmath $N_f$} hypermultiplets}

The only asymptotically free theories that we can consider within our approach,
for these groups, are $N_f=2,\ldots,2r$. Let us list some of the results that
we have obtained by using our formulas (we omit the case of $SU(N_c)$ whithout
matter which can be found in \cite{emm}):

\subsubsection{\boldmath $SU(2)$}

\noindent\underline{$N_f=2$}~
Using $u_2=a^2$, we found the following corrections:
\beqa
\F_1 &=& \rule{0mm}{10mm}\frac{u_2+m^2}{2u_2} ~,
\label{su2f1}\\
\F_2 &=& \rule{0mm}{10mm}\frac{u^2_2-6u_2 m^2 + 5 m^4}{64 u^3_2} ~,
\label{su2f2} \\
\F_3 &=& \rule{0mm}{10mm}\frac{5u^2_2 m^2-14u_2 m^4+9m^6}{192u_2^5} ~.
\label{su2f3}
\eeqa
The one- and two-instanton contributions coincide with those computed in
\cite{dhoker} (after adjusting $\Lambda$ to $\Lambda/2$).
In the one-instanton correction, as discussed before, it is possible to
split the masses, $m^2 \to m_1m_2$, so that the result for non-degenerated
$(ND)$ matter hypermultiplets is
\be
\F_{1,ND} = \rule{0mm}{10mm}\frac{u_2+m_1m_2}{2u_2} ~,
\label{su2non}
\ee
in agreement with the result in Ref.\cite{dhoker}.
Also, the case of one hypermultiplet can then be considered by letting
$m_2\to\infty$ while keeping $\Lambda m_2$ finite an equal to the square of
the dynamical scale $\Lambda^2$ that corresponds to $N_f=1$.

\subsubsection{\boldmath $SU(3)$}

To express our results, we introduce Weyl invariant combinations
in terms of the
$a_i$-variables, $u_2=a_1^2+a_2^2-a_1a_2$ and $u_3=a_1a_2 (a_1-a_2)$.
For the case $N_f=4$, we will denote $q_1=m_1+m_2$ and $q_2=m_1m_2$. We
obtained:

~

\noindent\underline{$N_f=2$}
\beqa
\F_1 &=&\rule{0mm}{10mm} {(2u_2^2+6m^2u_2-18mu_3)}/{\Delta_0} ~,
\label{su3f1} \\
\F_2 &=&\rule{0mm}{10mm}
(5u_2^6+153m^4u_2^4+162m^2u_2^5-1998m^3u_3u_2^3-414mu_3 u_2^4+1701m^4u_3^2u_2
\nonumber\\ &&+4374m^2u_3^2u_2^2+162u_3^2u_2^3+729u_3^4-2916m^3u_3^3-2673
mu_3^3u_2)/{\Delta_0^3} ~,
\label{su3f2} \\
\F_3 &=&\rule{0mm}{10mm}(48u_2^{10}+12320m^6u_2^7+31792m^4u_2^8+4992m^2u_2^9
-366624m^5u_3u_2^6-12032mu_3u_2^8\nonumber\\
&&-253088m^3u_3u_2^7+478116m^6u_3^2u_2^4+2276856m^4u_3^2
u_2^5+529236m^2u_3^2u_2^6\nonumber\\
&&+5600u_3^2u_2^7-3684852m^5u_3^3u_2^3-4654800m^3u_3^3u_2^4
-394524mu_3^3u_2^5+994356m^6u_3^4u_2\nonumber\\
&&+7097544m^4u_3^4u_2^2+3969648m^2u_3^4u_2^3+105192u_3^4u_2^4
-1469664m^5u_3^5\nonumber\\
&&-4878468m^3u_3^5u_2-1571724mu_3^5u_2^2+1364688m^2u_3^6+
215784u_3^6u_2)/{\Delta_0^{5}} ~.
\label{su3f3}
\eeqa
Again, the splitting of the one-instanton correction can be done by letting
$m^2 \to m_1m_2$ and $m \to (m_1+m_2)/2$,
\be
\F_{1,ND} = \rule{0mm}{8mm}(2u_2^2+6m_1m_2u_2-9(m_1+m_2)u_3)/{\Delta_0} ~,
\label{su3non1}
\ee
and the reduction of hypermultiplets mentioned above is immediate.
Our results (\ref{su3f1}), (\ref{su3f2}) and (\ref{su3non1}) agree with those
obtained in \cite{dhoker}.

~

\noindent\underline{$N_f=4$}
\be
\F_1 = \rule{0mm}{8mm}(2u_2^3-9u_3^2-6q_1u_2u_3+2(q_1^2+2q_2)u_2^2+6q_2^2u_2-
18q_1q_2u_3)/{\Delta_0} ~,
\label{su3bf1}
\ee
and the length of the expressions growths rapidly.
Again, having in mind that $\F_{1,ND}$ should be linear in the polynomials of
the masses it is possible to carefully split the masses, $m_1^2 \to m_1m_3$,
$m_2^{\,2} \to m_2m_4$, $m_1 \to (m_1+m_3)/2$ and $m_2 \to (m_2+m_4)/2$ (or, in
terms of the polynomials of the masses, $2q_1 \to t_1$, $q_1^2+2q_2 \to t_2$,
$2q_1q_2 \to t_3$ and $q_2^2 \to t_4$) so that
\be
\F_{1,ND} = \rule{0mm}{8mm}(2u_2^3-9u_3^2-3t_1u_2u_3+2t_2u_2^2-9t_3u_3
+6t_4u_2)/{\Delta_0} ~,
\label{su3non2}
\ee
and then one can reduce it to an odd number of matter hypermultiplets
in the way mentioned before \cite{dhoker}. 

\subsubsection{\boldmath $SU(4)$}

As the expressions become too long, we will only display the
one-instanton correction. To express our results,
 we introduce the classical
values of the Weyl invariant Casimirs 
$u_2=a_1^2+a_2^2+a_3^2-a_1a_2-a_2a_3$,
$u_3=a_1a_2(a_1-a_2)+a_2a_3(a_2-a_3)$ and $u_4=a_1^2a_2a_3-a_1a_2^2a_3
-a_1^2a_3^2+a_1a_2a_3^2$.
For the case $N_f=4$, we will denote $q_1=m_1+m_2$ and $q_2=m_1m_2$.
\vskip2mm
\noindent\underline{$N_f=2$}
\beqa
\F_1 &=& (8m^2u_2^3+6u_3^2u_2-8mu_3u_2^2-36u_3^2m^2+32m^2u_2u_4
\nonumber\\
&& -16u_2^2u_4+96mu_3u_4-64u_4^2)/{\Delta_0} ~.
\label{su4f1}
\eeqa
\noindent\underline{$N_f=4$}
\beqa
\F_1&=&(2u_3^2u_2^2-8u_4u_2^3+12u_3^2u_4-32u_4^2u_2-18q_1u_3^3
+64q_1u_4u_3u_2\nonumber\\
&&+6(q_1^2+2q_2)u_3^2u_2-16(q_1^2+2q_2)u_4u_2^2-64(q_1^2+2q_2)
u_4^2-8q_1q_2u_3u_2^2\nonumber\\
&&+96q_1q_2u_4u_3+8q_2^2u_2^3-36q_2^2u_3^2+32q_2^2u_4u_2)/
{\Delta_0} ~.
\label{su4bf1}
\eeqa

%%%%%%%%%%%%%%%%%%%%%%%%%%%%%%%%%%%%%%%%%%%%%%%%%%%%%%%%%%%%%%%%%%%%%%%%%%%%%
%%%%%%%%%%%%%%%%%%%%%%%%%%%%%%%%%%%%%%%%%%%%%%%%%%%%%%%%%%%%%%%%%%%%%%%%%%%%%

\subsection{{\boldmath $SO(2r)$} with {\boldmath $N_f$} hypermultiplets}

The only asymptotically free theories that we can consider within our
approach, for these groups, are $N_f=0,2,\ldots,2(r-2)$. Notice that the case
$N_f=0$ corresponds to vanishing values of $\xi^+$ and $\xi^-$ in
(\ref{elprep}), which in the formulas of the instanton corrections implies
$R=1$. Let us list some of the result one can easily obtain by using our
formulas:

\subsubsection{\boldmath $SO(4)$}

For this group the classical values of the Casimir operators in terms of the
$a_i$ are given by $u_2=2a_1^2+2a_2^2$, $u_4=-(a_1^4+a_2^4-2a_1^2a_2^2)$, and
we can only consider the pure case ($N_f=0$).

~

\noindent\underline{$N_f=0$}
\beqa
\F_1&=&\rule{0mm}{5mm}2^2u_2/{\Delta_0} ~, \\
\F_2&=&\rule{0mm}{10mm}2(5u_2^3-60u_2u_4)/{\Delta_0^3} ~, \\
\F_3&=&\rule{0mm}{10mm}2^5(3u_2^5-120u_2^3u_4+240u_4^2u_2)/{\Delta_0^5} ~.
\eeqa

\subsubsection{\boldmath $SO(6)$}

For this group we have $u_2=2a_1^2+2a_2^2+2a_3^2-2a_1a_2-2a_1a_3$,
$u_4=-(a_1^4+a_2^4+a_3^4-2a_1^3a_2-2a_1a_2^3-2a_1^3a_3-2a_1a_3^3+3a_1^2a_2^2
+3a_1^2a_3^2-2a_2^2a_3^2-2a_1^2a_2a_3+2a_1a_2^2a_3+2a_1a_2a_3^2)$ , 
$u_6=a_1^4a_2^2-2a_1^3a_2^3+a_1^2a_2^4-2a_1^4a_2a_3+2a_1^3a_2^2a_3+a_1^4a_3^2
+2a_1^3a_2a_3^2-2a_1^2a_2^2a_3^2-2a_1^3a_3^3+a_1^2a_3^4$,
and we can consider the cases $N_f=0$ and $N_f=2$.

~

\noindent\underline{$N_f=0$}
\beqa
\F_1&=&\rule{0mm}{5mm}2^2(-u_2u_4-9u_6)/{\Delta_0} ~,\\
\F_2&=&\rule{0mm}{10mm}2(-5u_2^5u_4^3-43u_2^3u_4^4-172u_2u_4^5-
60u_2^6u_4u_6-647u_2^4u_4^2u_6-1701u_2^2u_4^3u_6+36u_2^5u_6^2
\nonumber\\
&&-1827u_2^3u_4u_6^2-3276u_4^4
u_6+1323u_2u_4^2u_6^2-14337u_2^2u_6^3-21627u_4u_6^3)/{\Delta_0^3} ~.
\eeqa

\noindent\underline{$N_f=2$}
\be
\F_1=\rule{0mm}{5mm}2^2(-m^4u_2u_4-9m^4u_6-4m^2u_4^2+u_4^2u_2+12m^2u_2u_6
-4u_6u_2^2-3u_4u_6)/{\Delta_0} ~.
\ee
Again, the splitting of the masses is possible for the one-instanton
correction. This happens for all the classical groups.

\subsubsection{\boldmath $SO(8)$}

In this case the expressions of the Casimir operators in terms of the $a_i$
gets too long so we are not going to list them here. We have found the
following expressions:

\noindent\underline{$N_f=0$}
\be
\F_1=2^2(9u_2^3u_8-u_2^2u_4u_6+3u_2u_6^2+32u_2u_4u_8+
48u_6u_8-4u_4^2u_6)/{\Delta_0} ~.
\ee

\noindent\underline{$N_f=2$}
\beqa
\F_1&=&2^2(9m^4u_2^3u_8-m^4u_2^2u_4u_6+3m^4u_2u_6^2+32m^4u_2
u_4u_8+48m^4u_6u_8-4m^4u_4^2u_6\nonumber\\ &&
-4m^2u_6^2u_2^2-12m^2u_4u_6^2-u_2u_4u_6^2-9u_6^3+12m^2u_4u_8u_2^2+32m^2u_8u_4^2
+4u_8u_4^2u_2\nonumber\\
&&+16m^2u_8u_6u_2-3u_8u_6u_2^2+32u_8u_6u_4+128m^2u_8^2-48u_8^2u_2)/{\Delta_0}
~.
\eeqa
The one-instanton corrections agree with those computed in Ref.\cite{dhoker}.

%%%%%%%%%%%%%%%%%%%%%%%%%%%%%%%%%%%%%%%%%%%%%%%%%%%%%%%%%%%%%%%%%%%%%%%%%%%%%
%%%%%%%%%%%%%%%%%%%%%%%%%%%%%%%%%%%%%%%%%%%%%%%%%%%%%%%%%%%%%%%%%%%%%%%%%%%%%
%%%%%%%%%%%%%%%%%%%%%%%%%%%%%%%%%%%%%%%%%%%%%%%%%%%%%%%%%%%%%%%%%%%%%%%%%%%%%

\section{Results for non Simply-Laced Lie Algebras}
\setcounter{equation}{0}
\indent

\subsection{{\boldmath $SO(2r+1)$} with {\boldmath $N_f$} hypermultiplets}

In this case the form of $Z_{\alpha(G)}$ is different from the simply-laced
cases. Indeed, using the fact that
\[ \alpha_i\cdotsk\e_q = \delta_{i,q}-\delta_{i+1,q}
~~~~~~~~~~~ \alpha_r\cdotsk\e_q = 2\delta_{r,q} ~, \]
and setting $\xi^+=1$, we have
\beqa
Z_{\alpha(G)} &=& n^i\H_{2,i}^{(0)} ~= ~n^i\sum_pe_p\partial_ie_p
 \nonumber\\& = &n^i\sum_pe_p(\alpha_i\cdotsh\e_p)
= \sum_{i=1}^{r-1}~\left(e_i-e_{i+1}\right)n^i+2 e_r n^r
\nonumber\\ & = & Z_\alpha + Z_{\alpha_r} n^r  ~.
\label{zetagso}
\eeqa

The only asymptotically free theories that we can consider within our approach,
for $SO(2r+1)$, are $N_f=0,2,\ldots,2r-2$.
Notice that the case $N_f=0$ corresponds, as in $SO(2r)$,
to take in (\ref{elprep}) $\xi^+=\xi^-=0$ which in the formulas
of the instanton corrections means to set $R=1$. Let us list some
of the results that we have obtained:

\subsubsection{\boldmath $SO(5)$}

We can consider within our approach the cases $N_f=0$ and $N_f=2$. For this
group we have $u_2=2a_1^2+4a_2^2-4a_1a_2$,
$u_4=-(a_1^4-4a_1^3a_2+4a_1^2a_2^2)$. We found:

~

\noindent\underline{$N_f=0$}
\beqa
\F_1 &=& \rule{0mm}{5mm}-2^3u_4/{\Delta_0} ~,
\label{so5f1}
\\
\F_2 &=&\rule{0mm}{10mm} 2(u_2^3u_4^2-76u_2u_4^3)/{\Delta_0^3} ~,
\label{so5f2}
\\
\F_3 &=&
\rule{0mm}{10mm}2^{7}(3u_2^4u_4^4-232u_2^2u_4^5+176u_4^6)/{3\Delta_0^5} ~.
\label{so5f3}
\eeqa

\noindent\underline{$N_f=2$}
\beqa
\F_1 &=&\rule{0mm}{6mm} 2^2 (-2m^4u_4+2m^2u_2u_4-u_2^2u_4-2u_4^2)/{\Delta_0} ~,
\label{so5bf1}
\\
\F_2&=&\rule{0mm}{10mm}2(m^8u_2^3u_4^2-76m^8u_2u_4^3+152m^6u_2^2u_4^3
-32m^6u_4^4 -78m^4u_2^3u_4^3+168m^4u_2u_4^4\nonumber\\
&&+12m^2u_2^4u_4^3-88m^2u_2^2u_4^4+96m^2u_4^5-u_2^5u_4^3
-60u_2u_4^5+u_2^3u_4^4)/{\Delta_0^3} ~,
\label{so5bf2}
\\
\F_3 &=& \rule{0mm}{10mm}2^6(6m^{12}u_2^4u_4^4-464m^{12}u_2^2u_4^5
+352m^{12}u_4^6-9m^{10}u_2^5u_4^4+1256m^{10}u_2^3u_4^5\nonumber\\
&&-1744m^{10}u_2u_4^6+3504m^8u_2^2u_4^6+3m^8u_2^6u_4^4-1212m^8u_2^4u_4^5
-960m^8u_4^7\nonumber\\
&&+2976m^6u_2u_4^7-3024m^6u_2^3u_4^6+498m^6u_2^5u_4^5-2864m^4u_2^2u_4^7
+1054m^4u_2^4u_4^6\nonumber\\
&&-86m^4u_2^6u_4^5+736m^4u_4^8-1104m^2u_2u_4^8+824m^2u_2^3u_4^7
-137m^2u_2^5u_4^6\nonumber\\
&&+5m^2u_2^7u_4^5+240u_2^2u_4^8-128u_4^9-56u_2^4u_4^7+5u_2^6u_4^6)/{3\Delta_0^5}
~.
\eeqa

\subsubsection{\boldmath $SO(7)$}

Here we have $u_2=2a_1^2+2a_2^2+4a_3^2-2a_1a_2-4a_2a_3$~, $u_4=
-(a_1^4+a_2^4-2a_1^3a_2-2a_1a_2^3-4a_2^3a_3+3a_1^2a_2^2+8a_1^2a_3^2+4a_2^2a_3^2
-8a_1^2a_2a_3+8a_1a_2^2a_3-8a_1a_2a_3^2)$,
$u_6=a_1^4a_2^2+a_1^2a_2^4-2a_1^3a_2^3-4a_1^4a_2a_3+8a_1^3a_2^2a_3
-4a_1^2a_2^3a_3+4a_1^4a_3^2-8a_1^3a_2a_3^2+4a_1^2a_2^2a_3^2$~, and for $N_f=4$
we also denote $q_2=m_1^2+m_2^{2}$ and $q_4=m_1^2m_2^{\,2}$. We can consider
the cases $N_f=0$, $N_f=2$ and $N_f=4$.

~

\noindent\underline{$N_f=0$}
\beqa
\F_1&=&\rule{0mm}{5mm}2^3(u_2^2u_6+3u_4u_6)/{\Delta_0} ~,
\\
\F_2&=&\rule{0mm}{10mm}2\, u_6^2(-u_2^6u_4^3-12u_2^4u_4^4-48u_2^2u_4^5-
64u_4^6-76u_2^7u_4u_6\nonumber\\
&&-839u_2^5u_4^2u_6-3057u_2^3u_4^3u_6-3588u_2u_4^4u_6-44u_2^6u_6^2
-1695u_2^4u_4u_6^2\nonumber\\
&&-10827u_2^2u_4^2u_6^2-16308u_4^3u_6^2+1863u_2^3u_6^3-567u_2u_4u_6^3
-18225u_6^4)/{\Delta_0^3} ~.
\eeqa

\noindent\underline{$N_f=2$}
\be
\F_1=2^3u_6(m^4u_2^2+3m^4u_4+m^2u_2u_4+9m^2u_6-3u_2u_6)/{\Delta_0} ~.
\ee

\noindent\underline{$N_f=4$}
\beqa
\F_1&=&2^3u_6(q_4^2u_2^2+3q_4^2u_4-3(q_2^2+2q_4)u_2u_6
+(q_2^2+2q_4)u_4^2+q_2q_4u_2u_4+9q_2q_4u_6+3q_2u_4u_6\nonumber\\
&&+4q_2u_2^2u_6-q_2u_2u_4^2-45u_6^2-32u_2u_4u_6
-8u_2^3u_6+7u_4^3+2u_2^2u_4^2)/{\Delta_0} ~.
\eeqa
Again, the one-instanton corrections agree with previous results \cite{dhoker}.

%%%%%%%%%%%%%%%%%%%%%%%%%%%%%%%%%%%%%%%%%%%%%%%%%%%%%%%%%%%%%%%%%%%%%%%%%%%%%
%%%%%%%%%%%%%%%%%%%%%%%%%%%%%%%%%%%%%%%%%%%%%%%%%%%%%%%%%%%%%%%%%%%%%%%%%%%%%

\subsection{{\boldmath $Sp(2r)$} with {\boldmath $N_f$} hypermultiplets}

In this subsection we are going to consider the case of $Sp(2r)''$,
{\it i.e.}, the case of $Sp(2r)$ with two
massless hypermultiplets and $N_f'=N_f-2$ matter hypermultiplets.
In this case the form of $Z_{\alpha(G)}$ is, as in the case of $SO(2r+1)$,
different from the simply-laced cases. Now, using
\[ \alpha_i\cdotsk\e_q =  \delta_{i,q}-\delta_{i+1,q} 
~~~~~~~~~~~~~~~~ \alpha_r\cdotsk\e_q = \delta_{r,q} ~, \]
we see that
\beqa
Z_{\alpha(G)} &=& n^i\H_{2,i}^{(0)}\nonumber\\
 & \equiv & 2n^i\sum_pe_p\partial_ie_p
=2 n^i \sum_p e_p(\alpha_i\cdotsh\e_p)
= 2\left[\sum_{i=1}^{r-1}\left(e_i-e_{i+1}\right)n^i+  e_r
n^r\right] \nonumber\\ & = & 2 Z_{\alpha} -  Z_{\alpha_r}n^r  ~.
\label{zetagsp}
\eeqa

The only asymptotically free theories that we can consider within our approach,
for $Sp(2r)''$, are $N_f'=0,2,\ldots,2r-2$.
Notice that the case $N_f'=0$ now means to put $R(e_p)=e_p^4$ cause
we are considering two massless hypermultiplets. Let us list some
of the results that we have obtained:

\subsubsection{\boldmath $Sp(4)$}

For this group we can consider the cases $N_f'=0$ and $N_f'=2$. We have
$u_2=2a_1^2+a_2^2-2a_1a_2$, $u_4= -(a_1^4-2a_1^3a_2+a_1^2a_2^2)$.

~

\noindent\underline{$N_f'=0$}
\beqa
\F_1&=\rule{0mm}{5mm}&u_2/{2\Delta_0} ~,
\\
\F_2&=&\rule{0mm}{10mm}(5u_2^5+43u_2^3u_4+172u_2u_4^2)/4{\Delta_0^3} ~,
\\
\F_3&=&\rule{0mm}{10mm}2^2(9u_2^9+143u_2^7u_4+927u_2^5u_4^2+2840u_2^3u_4^3
+5680u_2u_4^4)/3{\Delta_0^5} ~.
\eeqa

\noindent\underline{$N_f'=2$}
\beqa
\F_1&=&\rule{0mm}{5mm}(m^4u_2+4m^2u_4-u_2u_4)/2{\Delta_0} ~,
\\
\F_2&=&\rule{0mm}{10mm}(5m^8u_2^5+43m^8u_2^3u_4+172m^8u_2u_4^2+12m^6u_2^4u_4
-24m^6u_2^2u_4^2+352m^6u_4^3+54m^4u_2^3u_4^2\nonumber\\
&&-264m^4u_2u_4^3+152m^2u_2^2u_4^3-32m^2u_4^4-5u_2^3u_4^3
+60u_2u_4^4)/4{\Delta_0^3} ~,
\\
\F_3&=&\rule{0mm}{10mm}2^2(9m^{12}u_2^9+143m^{12}u_2^7u_4+927m^{12}u_2^5u_4^2
+2840m^{12}u_2^3u_4^3\nonumber\\
&&+5680m^{12}u_2u_4^4+28m^{10}u_2^8u_4+304m^{10}u_2^6u_4^2+1536m^{10}u_2^4u_4^3
-1408m^{10}u_2^2u_4^4\nonumber\\
&&+9728m^{10}u_4^5-11952m^8u_2u_4^5+6600m^8u_2^3u_4^4+75m^8u_2^7u_4^2
+537m^8u_2^5u_4^3\nonumber\\
&&+15872m^6u_2^2u_4^5-1200m^6u_2^4u_4^4+120m^6u_2^6u_4^3-1792m^6u_4^6
+7760m^4u_2u_4^6\nonumber\\
&&+321m^4u_2^5u_4^4-5416m^4u_2^3u_4^5-3584m^2u_2^2u_4^6+1280m^2u_4^7
+752m^2u_2^4u_4^5\nonumber\\
&&-9u_2^5u_4^5+360u_2^3u_4^6-720u_2u_4^7)/3{\Delta_0^5} ~.
\eeqa

\subsubsection{\boldmath $Sp(6)$}

For this group we can consider the cases $N_f'=0$, $N_f'=2$ and $N_f'=4$.
We let $u_2=2a_1^2+2a_2^2+a_3^2-2a_1a_2-2a_2a_3$, $u_4=
-(a_1^4+a_2^4-2a_1^3a_2-2a_2^3a_3-2a_1a_2^3+3a_1^2a_2^2+2a_1^2a_3^2+a_2^2a_3^2
-4a_1^2a_2a_3+4a_1a_2^2a_3-2a_1a_2a_3^2)$ ,
$u_6=a_1^4a_2^2-2a_1^3a_2^3+a_1^2a_2^4-2a_1^4a_2a_3+4a_1^3a_2^2a_3
-2a_1^2a_2^3a_3+a_1^4a_3^2-2a_1^3a_2a_3^2+a_1^2a_2^2a_3^2$. For $N_f'=4$, and
we also have $q_2=m_1^2+m_2^{\,2}$ and
$q_4=m_1^2m_2^{2}$.

~

\noindent\underline{$N_f'=0$}
\beqa
\F_1&=&\rule{0mm}{5mm}2(3u_2u_6-u_2^2u_4-4u_4^2)/{\Delta_0} ~,
\\
\F_2&=&\rule{0mm}{10mm}2^4(-5u_2^6u_4^5-60u_2^4u_4^6-240u_2^2u_4^7-320u_4^8
+43u_2^7u_4^3u_6+546u_2^5u_4^4u_6+2304u_2^3u_4^5u_6\nonumber\\
&&+3232u_2u_4^6u_6-172u_2^8u_4u_6^2-2255u_2^6u_4^2u_6^2
-9715u_2^4u_4^3u_6^2-13272u_2^2u_4^4u_6^2+2064u_4^5u_6^2\nonumber\\
&&+180u_2^7u_6^3-1107u_2^5u_4u_6^3-18975u_2^3u_4^2u_6^3-46908u_2u_4^3u_6^3
+2439u_2^4u_6^4\nonumber\\
&&-3240u_2^2u_4u_6^4-57672u_4^2u_6^4+19197u_2u_6^5)/{\Delta_0^3} ~.
\eeqa

\noindent\underline{$N_f'=2$}
\beqa
\F_1=\rule{0mm}{7mm}2(3m^4u_2u_6-m^4u_2^2u_4-4m^4u_4^2-4m^2u_2^2u_6
-12m^2u_4u_6-u_2u_4u_6-9u_6^2)/{\Delta_0} ~.
\eeqa

\noindent\underline{$N_f'=4$}
\beqa
\F_1&=&\rule{0mm}{7mm}2(3q_4^2u_2u_6-q_4^2u_2^2u_4-4q_4^2u_4^2-4q_2q_4u_2^2u_6
-12q_2q_4u_4u_6-(q_2^2+2q_4) u_2u_4u_6\nonumber\\
&&-9(q_2^2+2q_4)u_6^2+12q_2u_2u_6^2-4q_2u_4^2u_6-3u_4u_6^2-4u_2^2u_6^2
+u_2u_4^2u_6)/{\Delta_0} ~.
\eeqa

%%%%%%%%%%%%%%%%%%%%%%%%%%%%%%%%%%%%%%%%%%%%%%%%%%%%%%%%%%%%%%%%%%%%%%%%%%%%%
%%%%%%%%%%%%%%%%%%%%%%%%%%%%%%%%%%%%%%%%%%%%%%%%%%%%%%%%%%%%%%%%%%%%%%%%%%%%%

\subsection{The case of pure {\boldmath $Sp(2r)$}}

As we discussed above, the case of $Sp(2r)$ without matter hypermultiplets
cannot be obtained from our previous formulas, as long as we are
considering at least two massless hypermultiplets.
Nevertheless, one can treat this case separately in an analogous way.
In fact, we can fix our ansatz for the effective prepotential (\ref{elprep}) to
the one first considered by Ito and Sasakura \cite{itosak} by setting
\be
\xi=1~~~~~~~~~\xi^+=\xi^-=0 ~~~~~~~~\mbox{and}~~~~~~~~~~ \tau_0 = 3 ~.
\label{settings}
\ee
Now, we can introduce the effective prepotential into Eq.(\ref{formula}) and
the same kind of formulas for the instanton correction would be obtained,
provided we have for this case a characteristic $E=\left[{0,\dots,0\atop
\med,\dots,\med}\right]$. Note that, being $N_f=0$, we must set $R=1$ in our
formulas.

We also need the value of $Z_{\alpha(G)}$ which turns out to be the same
as that in $Sp(2r)''$, \ie \ $Z_{\alpha(G)}=2Z_\alpha-n_rZ_{\alpha_r}$.

\subsubsection{\boldmath $Sp(4)$}

For this group we have, as we saw before, $u_2=2a_1^2+a_2^2-2a_1a_2$, $u_4=
-(a_1^4-2a_1^3a_2+a_1^2a_2^2)$. In terms of them the first instanton
corrections are

\beqa
\F_1&=&\rule{0mm}{5mm}2^3(u_2^2+4u_4)/{\Delta_0} ~,
\\
\F_2&=&\rule{0mm}{10mm}2^6(5u_2^7+59u_2^5u_4+232u_2^3u_4^2
+304u_2u_4^3)/{\Delta_0^3} ~,
\\
\F_3&=&\rule{0mm}{10mm}2^{14}(9u_2^{12}+184u_2^{10}u_4+1526u_2^8u_4^2
+6496u_2^6u_4^3+14656u_2^4u_4^4\nonumber\\
&&+15872u_2^2u_4^5+5632u_4^6)/{3\Delta_0^5} ~.
\eeqa

\subsubsection{\boldmath $Sp(6)$}

For this group we have $u_2=2a_1^2+2a_2^2+a_3^2-2a_1a_2-2a_2a_3$, $u_4=
-(a_1^4+a_2^4-2a_1^3a_2-2a_2^3a_3-2a_1a_2^3+3a_1^2a_2^2+2a_1^2a_3^2+a_2^2a_3^2
-4a_1^2a_2a_3+4a_1a_2^2a_3-2a_1a_2a_3^2)$ ,
$u_6=a_1^4a_2^2-2a_1^3a_2^3+a_1^2a_2^4-2a_1^4a_2a_3+4a_1^3a_2^2a_3
-2a_1^2a_2^3a_3+a_1^4a_3^2-2a_1^3a_2a_3^2+a_1^2a_2^2a_3^2$, and the first
instanton corrections are

\beqa
\F_1&=&\rule{0mm}{5mm}2^5(u_2^2u_4^2+4u_4^3-4u_2^3u_6-18u_2u_4u_6
-27u_6^2)/{\Delta_0} ~,
\\
\F_2&=&\rule{0mm}{10mm}
2^{12}(-5u_2^6u_4^7-60u_2^4u_4^8-240u_2^2u_4^9-320u_4^{10}+59u_2^7u_4^5u_6
+738u_2^5u_4^6u_6\nonumber\\
&&+3072u_2^3u_4^7u_6+4256u_2u_4^8u_6-232u_2^8u_4^3u_6^2-3021u_2^6u_4^4u_6^2
-12699u_2^4u_4^5u_6^2\nonumber\\
&&-15736u_2^2u_4^6u_6^2+6288u_4^7u_6^2+304u_2^9u_4u_6^3+4120u_2^7u_4^2u_6^3
+15518u_2^5u_4^3u_6^3+1716u_2^3u_4^4u_6^3\nonumber\\
&&-55728u_2u_4^5u_6^3-16u_2^8u_6^4+5928u_2^6u_4u_6^4+54486u_2^4u_4^2u_6^4
+113373u_2^2u_4^3u_6^4-216u_2^5u_6^5\nonumber\\
&&-41148u_4^4u_6^4+39447u_2^3u_4u_6^5+182250u_2u_4^2u_6^5-729u_2^2u_6^6
+89667u_4u_6^6)/{\Delta_0^3} ~.
\eeqa
The one-instanton corrections agree with those computed in Ref.\cite{itosak}.

%%%%%%%%%%%%%%%%%%%%%%%%%%%%%%%%%%%%%%%%%%%%%%%%%%%%%%%%%%%%%%%%%%%%%%%%%%%%%
%%%%%%%%%%%%%%%%%%%%%%%%%%%%%%%%%%%%%%%%%%%%%%%%%%%%%%%%%%%%%%%%%%%%%%%%%%%%%
%%%%%%%%%%%%%%%%%%%%%%%%%%%%%%%%%%%%%%%%%%%%%%%%%%%%%%%%%%%%%%%%%%%%%%%%%%%%%

\section{Concluding Remarks}
\setcounter{equation}{0}
\indent

In the present paper, we have shown how instanton corrections to the effective
prepotential of ${\cal N}=2$ supersymmetric theories can be computed in a
variety of cases including all classical gauge groups and even number of
degenerated fundamental matter hypermultiplets, {\em up to arbitrary order}.
As compared to other approaches developed in the literature, we should stress
that the one presented in this paper has an important feature in that it does
not require an explicit knowledge of the BPS spectrum as a function of the
moduli, at the same time that it allows to consider a huge variety of
cases within a unified framework.
Also, being recursive, it admits an easy implementation on a computer.
We have illustrated the remarkable simplicity of our procedure by displaying
many explicit expressions which should be quite useful for further comparison
with the results obtained by other means.

Conversely, our results admit a second reading: They could be thought of as a
highly non-trivial test of the connection between the Seiberg--Witten solution
to the low energy dynamics of ${\cal N}=2$ supersymmetric gauge theories, and
the theory of Whitham (adiabatic) deformations of a given integrable system
\cite{emm,ITEP}. In this sense, it is important to remind that the {\em new}
equation (\ref{formula}), which is a key ingredient of our procedure, is
originated in the latter framework, as it is shown in detail in
Ref.\cite{egrmm}.

Aside from being an interesting mathematical problem by itself, the embedding
of the Seiberg-Witten solution within a Whitham hierarchy seems to be the
appropriate framework for the study of many physical phenomena.
For example, the so-called {\em slow Whitham times} can be consistently
thought of as spurion vector supermultiplets that can be used to break ${\cal
N}=2$ supersymmetry down to ${\cal N}=0$ with non-quadratic Casimir operators
\cite{emm}.
In this way, the Whitham hierarchy can be interpreted as a family of
supersymmetry breaking deformations of the original theory associated with the
higher Casimir operators of the gauge group.
This issue generalizes to the ${\cal N}=0$ case the family of ${\cal N}=1$
supersymmetry breaking terms considered, for instance, in Ref.\cite{ad}.

The key feature of the Whitham formalism lies on the fact that, as the
dependence on the slow times is encoded in the prepotential, it is possible to
obtain the {\em exact} effective potential of the theory, in the spirit of
\cite{emm,soft}.
This allows to perform a detailed study, both qualitative and quantitative,
of the vacuum state of the theory once supersymmetry is broken, as well as
of the appearance of monopole condensates, mass gaps, etc.
Preliminary results on this program were published in Ref.\cite{emm}. The
formalism is even useful near the Argyres--Douglas singularities, where
non-local degrees of freedom become simultaneously massless, provided one
approaches them along any of the submanifolds where a unique monopole gets
massless \cite{emm}.

There is another place where deformations of the prepotential by means of the
Whitham times are relevant. It is the study of contact terms in the twisted
version of ${\cal N}=2$ gauge theories, where these new variables play the
r\^ole of sources for insertions of certain class of operators in the
generating functional \cite{marimoore,losev,wm,takasaki}. 

Another interesting point is given by the uses of our starting equations
(\ref{lderm})--(\ref{formula}) to study the strong coupling expansion of the
prepotential near the singularities of the quantum moduli space, as it was
done in Ref.\cite{edemas} for the case of pure $SU(N_c)$. 
In particular, these equations provide us with a set of non-trivial
constraints that facilitate the study of the couplings between different {\em
magnetic photons}, originally found in Ref.\cite{ds}, that take place at such
points. The expansion of the prepotential near the maximal points by other
methods, as the deformations of the auxiliary singular Riemann manifold
\cite{dHP}, is not sensitive to such kind of terms.

Several interesting questions remain open aside from the ones just mentioned.
For example, the case of arbitrary masses cannot be treated within our
approach, except for the one-instanton correction (which, on the other hand,
is enough for leading order comparison purposes). In fact, from the Whitham
hierarchy side, one can show that indeed the formulas used in this paper are
insufficient to tackle the generic scenario, though it seems to be possible to
refine the formalism in order to extend its applicability to some cases of
unpaired masses \cite{egrmm}. The additional corrections that appear in the
generic case are, nevertheless, quite difficult to manage with. Finally, 
another avenue for future research is, certainly, the connection of this
formalism with the string theory and D-brane approach to supersymmetric gauge
theories, where some steps has already been given in the last few years
\cite{gorsky}.

We believe that these matters deserve further study.

%%%%%%%%%%%%%%%%%%%%%%%%%%%%%%%%%%%%%%%%%%%%%%%%%%%%%%%%%%%%%%%%%%%%%%%%%%%%%
%%%%%%%%%%%%%%%%%%%%%%%%%%%%%%%%%%%%%%%%%%%%%%%%%%%%%%%%%%%%%%%%%%%%%%%%%%%%%
%%%%%%%%%%%%%%%%%%%%%%%%%%%%%%%%%%%%%%%%%%%%%%%%%%%%%%%%%%%%%%%%%%%%%%%%%%%%%

\section*{Acknowledgements}

We are pleased to thank Marcos Mari\~no for helpful comments and a careful
reading of the manuscript. The work of J.D.E. is supported by a fellowship of
the Ministry of Education and Culture of Spain.  The work of J.M. was
partially supported by DGCIYT under contract PB96-0960.

%%%%%%%%%%%%%%%%%%%%%%%%%%%%%%%%%%%%%%%%%%%%%%%%%%%%%%%%%%%%%%%%%%%%%%%%%%%%%
%%%%%%%%%%%%%%%%%%%%%%%%%%%%%%%%%%%%%%%%%%%%%%%%%%%%%%%%%%%%%%%%%%%%%%%%%%%%%
%%%%%%%%%%%%%%%%%%%%%%%%%%%%%%%%%%%%%%%%%%%%%%%%%%%%%%%%%%%%%%%%%%%%%%%%%%%%%

%---------------- Bibliografia-------------------

\end{document}